# Ion Beam radiation and temperature effect on Co/Si and CoO/Co/Si thin films

A.S.Bhattacharyya

Centre for Nanotechnology
Central University of Jharkhand
Brambe, Ranchi: 835205

Email: 2006asb@gmail.com

**Abstract**
Co and CoO thin films were deposited by magnetron sputtering in the form of multilayers. They were irradiated with Ar and Si ion beams of different energies and fluences. The magnetic properties were investigated.

**Keywords:** Co, CoO, thin films, Ion beam, magnetic properties

Ion beam irradiation is an effective means of studying the morphological changes occurring in thin films. Surface energy changes as well as displacement of atoms due to localized heating can bring about interesting phenomena which can lead to possible patterning and device fabrication in the future [1]. Ion beam irradiation was performed previously on Co/CoO films deposited on Si (111) substrates by magnetron sputtering. Ion-beam-induced dewetting and exchange bias was observed [2]. This communication is an extension of the previous work [2].

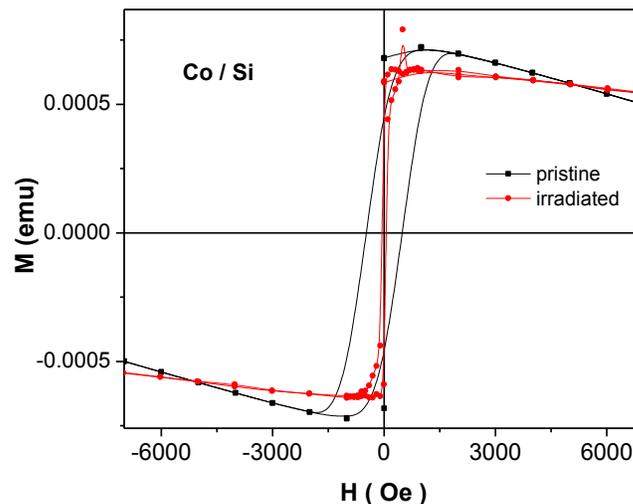

**Fig 1** : MH plots for Si ions irradiation on Co/Si thin films

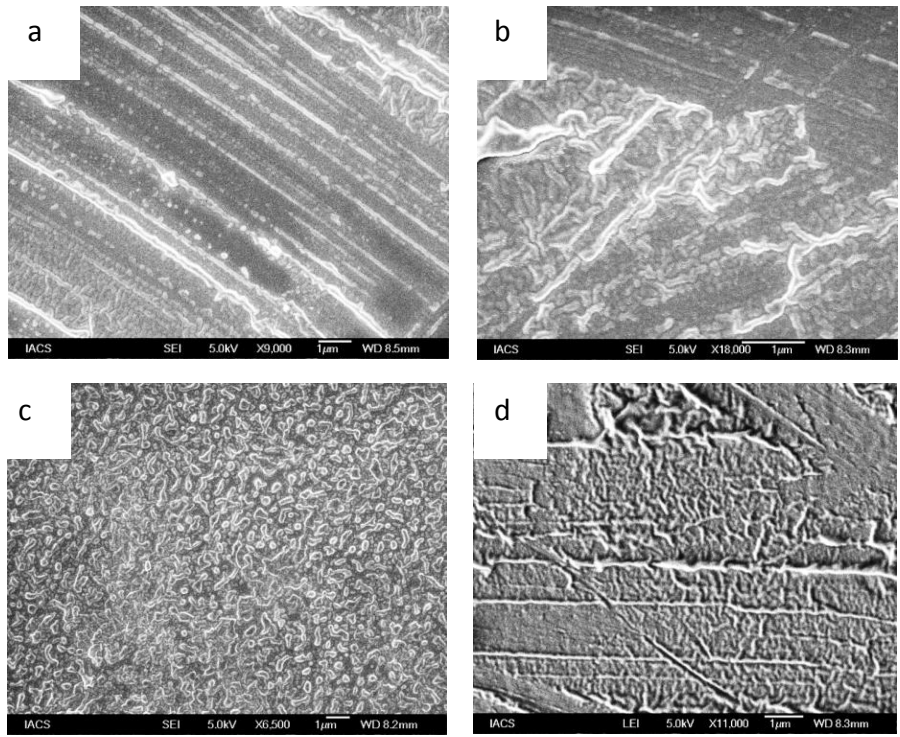

**Fig 2**: SEM of Cobalt thin films

Thickness of each film was 25 nm. The films were then irradiated with Si and Ar ions at different fluences and energies. Scanning electron microscopy (SEM) and magnetic force microscopy (MFM) measurements were performed to study the changes in morphology and magnetic structure that occurred with the irradiation while SQUID and PPMS measurements were performed to study temperature and magnetic field dependence of dc magnetization. Superparamagnetism was observed for Si ion irradiated Co films on Si(111) substrates as shown in Fig 1. SEM of the deposited and irradiated films are shown in Fig 2. Ion beam induced dewetting as proposed by lian et al can be observed in fig 2 c. Grooves can be seen on the surface which is due to enhancement of irregularities at the surface due to preferential etching [3].

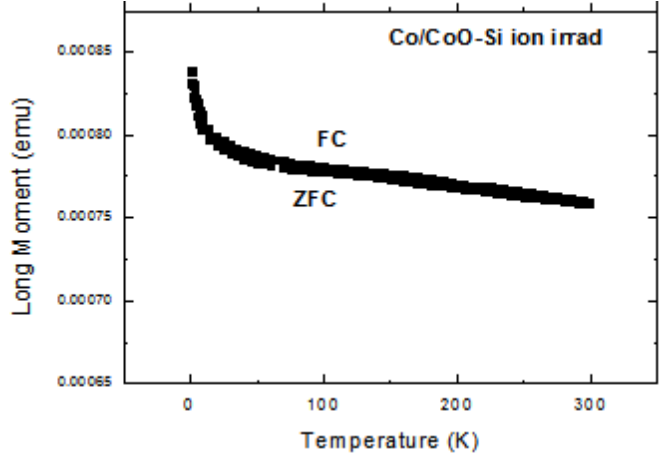

Fig 3: ZFC-FC magnetization curves for Si ion Fluence of $10^{17}$ with energy 50 KeV on Co/CoO thin films.

Superparamagnetism was observed for Co/CoO films, which were irradiated with 200 keV Ar ions at a fluence of $10^{15}$ ions/cm². Blocking temperature around 300 K was obtained from the measured ZFC-FC magnetization curves which has been published elsewhere [3]. An exchange bias was also observed with variation Ai ion fluence from $10^{15}$ to $10^{16}$ ions/cm² (Fig 4). A Si ion Fluence of $10^{17}$ with energy 50 KeV however did not show any deviation in the ZFC-FC magnetization curves (Fig 3). Interestingly although an increase in coercivity and exchange bias was observed previously for ion beam irradiated Co/CoO films, a decrease in coercivity was observed in pristine Co/Si with temperature as shown in Fig 3. Effect of temperature on the magnetic properties is a complex area and there are numerously results. In the case of Ni ferrite nanoparticles, an increase in coercivity with decrease in temperature has been observed which was explained on the basis of presence of freezed surface-spins and some paramagnetic impurities at the shell of nanoparticles that are activated at lower temperatures [7]. These grooves if properly controlled and patterned can act as enhanced optical transmitting zones [8].

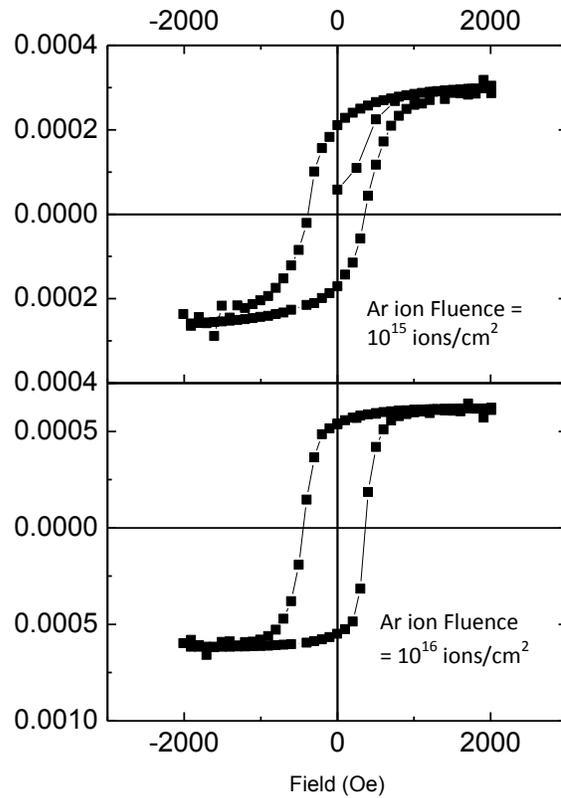

Fig 4: Exchange bias with variation Ai ion fluence from $10^{15}$ to $10^{16}$ ions/cm²

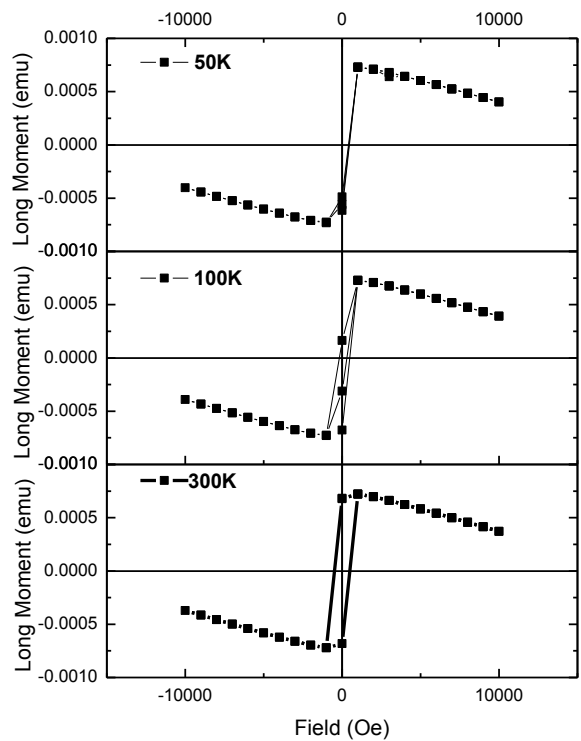

Fig 5: A decrease in coercivity was observed in pristine Co/Si with temperature

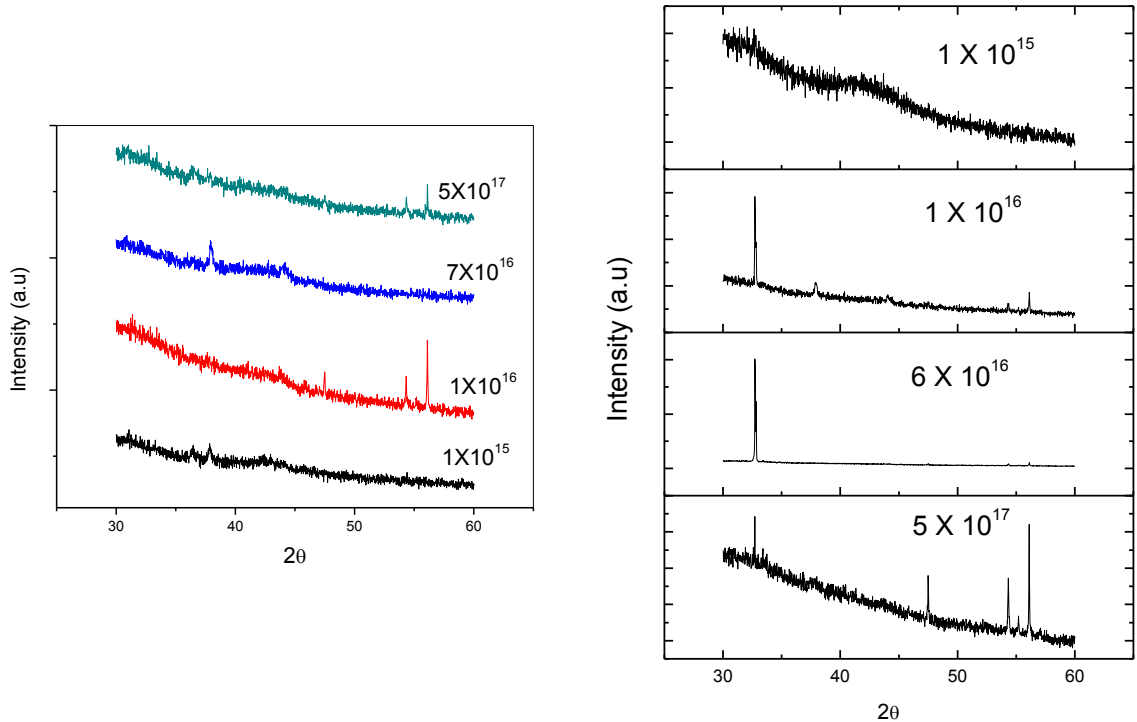

**Fig 6 :** XRD of Co/Si films irradiated at different fluences

**ACKNOWLEDGMENTS**

The authors acknowledge Prof B.N.Dev, IACS Kolkata, Prof S.M.Yusuf, BARC Mumbai, Dr. D.Kabiraj, IUAC New Delhi and D.P. Mahapatra, IOP Bhubaneswar, for help with ion beam irradiation.